\documentclass{emulateapj}
\usepackage{apjfonts}

\newcommand{\syn}{\rm syn}

\shorttitle{Evolution of Synchrotron X-rays in SNRs}
\shortauthors{Nakamura et al.}

\begin{document}


\title{Evolution of Synchrotron X-rays in Supernova Remnants}

\author{
Ryoko Nakamura\altaffilmark{1,2},
Aya Bamba\altaffilmark{3,1,4},
Tadayasu Dotani\altaffilmark{1},
Manabu Ishida\altaffilmark{1},
Ryo Yamazaki\altaffilmark{4},
Kazunori Kohri\altaffilmark{5}
}

\altaffiltext{1}{
ISAS/JAXA Department of High Energy Astrophysics
3-1-1 Yoshinodai, Chuo-ku, Sagamihara,
Kanagawa 252-5210, Japan
}

\altaffiltext{2}{
Department of Physics, Tokyo Institute of Technology, 2-12-1 Ookayama,
Meguro-ku, Tokyo 152-8551, Japan}

\altaffiltext{3}{
School of Cosmic Physics, Dublin Institute for Advanced Studies
31 Fitzwilliam Place, Dublin 2,
Ireland
}

\altaffiltext{4}{
Department of Physics and Mathematics,
Aoyama-Gakuin University
5-10-1 Fuchinobe, Sagamihara, Kanagawa, 252-5258, Japan
}

\altaffiltext{5}{
Theory Center, Institute of Particle and Nuclear Studies,
KEK (High Energy Accelerator Research Organization),
1-1 Oho, Tsukuba 305-0801, Japan
}

\begin{abstract}
A systematic study of 
the synchrotron X-ray emission from supernova remnants (SNRs)
has been conducted.
We selected a total of 12 SNRs
whose synchrotron X-ray spectral parameters are available
in the literature with reasonable accuracy,
and studied how their luminosities change
as a function of radius.
It is found that the
synchrotron X-ray luminosity tends to drop 
especially when the SNRs become larger than $\sim$5~pc,
despite large scatter.
This may be explained by the change of
spectral shape caused by
the decrease of the synchrotron roll-off energy.
A simple evolutionary model
of the X-ray luminosity is proposed
and is found to reproduce the observed data approximately,
with reasonable model parameters.
According to the model,
the total energy of accelerated electrons is
estimated to be $10^{\rm 47-48}$~ergs,
which is well below the supernova explosion energy.
The maximum energies of accelerated electrons and protons
are also discussed.
\end{abstract}


\section{Introduction}

Young supernova remnants (SNRs) are widely believed to be
the main source of Galactic cosmic rays.
\citet{koyama1995} discovered synchrotron X-rays
from shells of SN~1006,
which was the first observational clue of 
cosmic-ray electrons being accelerated up to the TeV energy range.
Later, several young SNRs were found to have synchrotron X-ray shells
\citep[c.f.,][]{koyama1997,slane1999,bamba2000,vink2003}.
Another piece of evidence for cosmic-ray acceleration in SNRs 
was obtained from gamma-ray observations.
Very-high-energy (VHE) and GeV gamma-rays have been detected 
from several SNRs,
although, it is still unclear whether their origin
is hadronic or leptonic
\citep[e.g.,][]{aharonian2004,aharonian2007,
abdo2010b,abdo2010c,abdo2011}.

Despite these various pieces of evidence
in favor of acceleration,
it is still unclear how the acceleration process
evolves in SNRs,
in particular how the accelerated particles cool down and
how they escape from SNRs.
One reason for this lack of understanding is that
past observational studies have concentrated on
individual sources alone.
In this paper, we investigate, for the first time,
the time evolution of SNR synchrotron X-rays,
using all available data in literature for sources
in which this emission component is bright.
In section~\ref{sec:sample},
we describe the evolution of synchrotron X-ray luminosity
in our sample.
An interpretation and a simple model to support it
are in section~\ref{sec:model}.
Finally, section~\ref{sec:discuss} is devoted to
discussion of the results.

\section{Sample selection and Results}
\label{sec:sample}

We searched the literature for reports of synchrotron X-ray emission
from SNRs and found a total of 14 such sources.
However, we could not use all of these for the current study,
for various reasons.
Kepler \citep{bamba2005,reynolds2007}
and G330.2+1.0 \citep{park2009} had to be removed from our sample,
because of the difficulty of estimating
the total synchrotron X-ray luminosity,
caused by a low detection significance or
by contamination from thermal X-rays.
Thus our sample consists of the 12 SNRs
listed in Table~\ref{tab:sample}.
We used the latest value of the 2--10~keV synchrotron X-ray unabsorbed flux
from the references listed in the table.
When the reported flux was not absorption-corrected or
was for a different energy band,
we normalized it unabsorbed flux in the 2--10~keV
using the best-fit model in the references.
Synchrotron X-ray emission may have a spectral roll-off
in the 2--10~keV band.
When significant roll-off was reported in the references,
we took such a component into account while computing the flux.
Some SNRs (Cas~A, Tycho, SN~1006)
have bright thermal X-ray emission,
which makes it difficult to estimate the synchrotron X-ray flux.
In these cases,
we inferred the synchrotron X-ray flux
based on wide-band information and/or detailed spectroscopy 
to isolate synchrotron X-rays from thermal component,
as we believe such analysis can provide the most reliable 
results currently available.
Two of the sample sources, W28 and G156.7+5.7, are largely extended but
were only partly observed with enough exposure in X-rays.
As the X-ray luminosities can be deduced only for 
the observed regions
(5\% for W28 and 26\% for G156.7+5.7, respectively),
possible contributions from the unobserved regions
are included in the upper error bars.
When we estimated these contributions,
we assumed that the surface brightness of the unobserved region
in each source is identical to that of the observed one.
The distance uncertainty is also included into the errors
on the luminosity.
We used a 10\% distance uncertainty 
for cases where 
no uncertainty was available in the literature.
In summary, we took into account the following uncertainties;
statistical errors given in literatures, systematic distance uncertainty,
and the limited coverage of the SNR extension.

In order to study the time evolution, we need to know
the SNR ages.
However, we have only four SNRs whose ages are historically known.
Another parameter, the ionization time scale of heated plasma,
is also difficult to determine
since several SNRs show no thermal emission.
We thus use the physical radius of each SNR, $R_s$, as the age indicator.
$R_s$ is taken from \citet{green2009}.
The distance uncertainty is included in the error computation.
Some SNRs have distorted shapes,
as shown in Tab.~\ref{tab:sample}.
This is also included in the errors on the radii.
All the derived parameters are shown in Tab.~\ref{tab:sample}.
The radius of a SNR depends on the density of the interstellar medium,
but rather insensitively.
For remnants in the Sedov phase, $R_s\propto n_0{}^{-1/5}$
(where $n_0$ is the upstream number density).
The radius changes by only a factor of $\sim$2 
even if the upstream density changes by 2~orders of magnitude.
The radius also has very weak dependency on the explosion energy $E_0$,
$R_s\propto E_0^{1/5}$.
On the other hand, $R_s\propto t_{\rm age}{}^{2/5}$
(where $t_{\rm age}$ is the age of the remnant).
Thus the radius is a good age indicator.

\begin{figure}
\epsscale{0.7}
\plotone{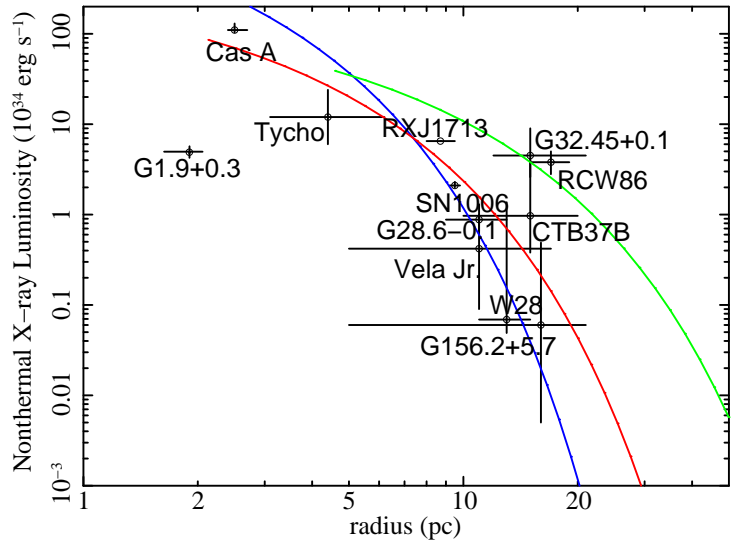}
\caption{
Synchrotron X-ray luminosity in 2--10~keV band as a function of
 radius for Galactic SNRs.
Solid lines represent
the evolutionary model of synchrotron X-rays
for $n_0$ of 5.0~cm$^{-3}$ (blue), 1.0~cm$^{-3}$ (red),
and 0.1~cm$^{-3}$ (green).
We plot results only for the Sedov phase ($t_1<t_{\rm age}<t_2$),
in which the normalization of the electron energy distribution, $A$, 
is expected to be constant with time (see text for details).
}
\label{fig:r_Lx}
\end{figure}

Figure~\ref{fig:r_Lx} shows the synchrotron X-ray luminosity
as a function of the radius.
One can see that
the luminosity is in the order of $10^{34}$--$10^{36}$~ergs~s$^{-1}$
when the SNRs are smaller than $R_s\sim$5~pc,
which corresponds to an age of a few hundred years,
whereas it decreses to $10^{32}$--$10^{35}$~ergs~s$^{-1}$
beyond $R_s\sim$5~pc.
it appears that the luminosity drops off rapidly
at $R_s\sim$5~pc,
although the scatter is rather large.
Note that all the SNRs with $R_s<5$~pc display
synchrotron X-rays brighter than $10^{34}$~ergs~s$^{-1}$ \citep{green2009},
whereas most of larger SNRs do not have significant synchrotron X-rays.
This fact makes the difference of luminosities larger
between the two regions.
The drop off in non-thermal X-ray luminosity reaches two or three
orders of magnitude,
which is much larger than the errors on the individual luminosities.
We have carried out a similar analysis for the thermal X-ray luminosity
for SNRs in the Large Magellanic Cloud \citep{williams1999}
and the radio luminosity of Galactic SNRs \citep{green2009,case1998}.
However, no drop off like that observed for synchrotron X-rays
has been found.

\section{Evolution of Synchrotron X-rays}
\label{sec:model}

Here we consider the cause of the decrease in synchrotron X-ray emission
identified in the previous section.
The $\nu F_\nu$-spectrum of synchrotron X-ray radiation 
has a peak around the roll-off frequency, $\nu_{\rm roll}$,
above which the flux rapidly decays towards higher frequencies.
As SNRs evolve, $\nu_{\rm roll}$ is known to decrease
\citep[e.g.,][]{bamba2005}.
The observed rapid decay in synchrotron X-rays around 
$R_s\sim10~{\rm pc}$ could be caused by
$\nu_{\rm roll}$ passing through the X-ray band
to lower energies.
In this section, we discuss this possibility in detail.
We introduce essential argument first,
and show a simple model to support it later.

We assume that the electron acceleration is energy-loss-limited,
in which the maximum energy of electrons, $E_e{}^{\max}$,
is determined from 
the balance of the synchrotron loss and acceleration:
\begin{equation}
E_e{}^{max} = \frac{24}{\xi^{1/2}}\left(\frac{v_s}{10^8~{\rm cm~s^{-1}}}\right)
\left(\frac{B_d}{10~\mu{\rm G}}\right)^{-1/2}\ \ {\rm (TeV)}\ \ ,
\label{eq:Emax_e}
\end{equation}
where $\xi$ is the gyro-factor.
%
In deriving Eq.~(\ref{eq:Emax_e}), we equate the acceleration time,
$t_{\rm acc}(E)=20\xi cE/3ev_s{}^2B_d$ with the synchrotron cooling time, 
$t_{\syn}(E)=125~{\rm yr}(E/10~{\rm TeV})^{-1}(B_d/100~\mu{\rm G})^{-2}$,
where $v_s$ and $B_d$ are the shock velocity and the downstream
magnetic field, respectively.
In this case, we derive 
\begin{equation}
h\nu_{\rm roll}\sim0.4~{\rm keV}~
\xi^{-1}(v_s/10^8~{\rm cm}~{\rm s}^{-1})^2 ~~,
\label{eq:rolloff}
\end{equation}
which is independent of $B_d$ \citep[e.g.,][]{aharonian1999,yamazaki2006}.
The roll-off frequency and the shock velocity have been measured
in several SNRs
\citep[Cas~A, SN~1006;][]{reynolds1999,patnaude2009,bamba2008,ghavamian2002}.
For these SNRs, eq.(\ref{eq:rolloff}) is consistent
with the observational values
to within 1 order of magnitude
assuming $\xi = 1$.
Also note that the effect of particle escape from the shock region
\citep[e.g.,][]{reynolds1998,ohira2010} is not considered in this paper.
This effect might be important for older SNRs
\citep{ohira2011b}.
However, our present model reproduces the observed trend well,
suggesting that particle escape is not yet significant for these SNRs
(see also the last paragraph in section~4).

The X-ray spectrum of synchrotron radiation is well 
approximated analytically.
It is mainly determined by shock dynamics in the SNR
if the synchrotron X-ray-emitting electrons with energy $E$ satisfy
$t_{\syn}(E)<t_{\rm age}$, where $t_{\rm age}$ is the age of the SNR.
In this case, the energy spectrum of electrons is generally given by
\begin{equation}
N_e(E) = AE^{-p}(1+E/E_b)^{-1}
\exp[-(E/E_{e}{}^{max})^2]\ \ ,
\label{eq:N_e}
\end{equation}
where the break energy, $E_b$,  is determined by 
$t_{\syn}(E_b)=t_{\rm age}$.
During the acceleration, electrons with $E>E_b$ suffer energy loss via
synchrotron cooling, which causes a steepening of the
energy spectrum \citep[e.g.,][]{longair94}.
We also note that the shape of the cutoff is analytically obtained in
\citet{zirakashvilli2007}.
%
%
Given the electron distribution, we calculate the approximate
formula of the X-ray luminosity, $L_\nu$.
The characteristic frequency, $\nu_{\syn}(E)$, of the synchrotron radiation
emitted by electrons with energy $E$ is given by
\begin{equation}
h\nu_{\syn}(E)\sim
0.12~{\rm keV}(B_d/10~\mu{\rm G})(E/10~{\rm TeV})^2~~.
\label{eq:character}
\end{equation}
Then, as long as $\nu<\nu_b=\nu_{\syn}(E_b)$,
we can apply the standard formula,
$L_\nu\propto AB_d^{(p+1)/2}\nu^{-(p-1)/2}$
\citep[e.g.,][]{longair94}.
On the other hand, if $\nu_b<\nu$, the spectral slope
steepens ($L_\nu\propto\nu^{-p/2}$) due to the 
steepening of the electron distribution, and we derive
\begin{eqnarray}
L_\nu &\propto& AB_d{}^{(p+1)/2}\nu_b{}^{-(p-1)/2} (\nu/\nu_b)^{-p/2}
\exp(-\sqrt{\nu/\nu_{\rm roll}})
\nonumber \\
&\propto&
AB_d{}^{(p-2)/2}\nu^{-p/2} \exp(-\sqrt{\nu/\nu_{\rm roll}})
~~,
\label{eq:Luminoosity}
\end{eqnarray}
where we assume that $L_\nu$ is continuous at $\nu = \nu_b$,
and we again use the result of \citet{zirakashvilli2007}
for the cutoff shape.
Calculating $\nu_b$ as
\begin{equation}
h\nu_b \sim 0.19~{\rm keV}
(B_d/10~\mu{\rm G})^{-3}(t_{\rm age}/10^4{\rm yr})^{-2}~~,
\end{equation}
we find that throughout the evolution of the SNR,
the X-ray band (2--10~keV) always lies above $\nu_b$ because
for young SNRs ($t_{\rm age}\lesssim10^3$~yr),
the magnetic field may be amplified to $B_d\gg10~\mu$G
\citep[e.g.,][]{bamba2003,bamba2005,vink2003}.
Equation.~(\ref{eq:Luminoosity}) is thus a good approximation
of the X-ray luminosity for any arbitrary epoch.
In particular, when $p\approx2$, the X-ray luminosity
is insensitive to the magnetic field,
depending instead mainly on $\nu_{\rm roll}$.
As the SNR ages, the shock velocity $v_s$ decreases and
$\nu_{\rm roll}$ becomes smaller. 
Equation~(\ref{eq:rolloff}) tells us that
$\nu_{\rm roll}$ is below the X-ray band (2--10~keV)
when $v_s\lesssim10^8$cm~s$^{-1}$, 
so that the X-ray luminosity drops off.

In order to demonstrate the above argument,
we construct a simple model
to calculate the synchrotron X-ray flux.
In our model, a simple shock dynamics scenario is considered.
We assume that the forward shock velocity of SNRs $v_s$ 
is a function of the age of SNR $t_{age}$ as follows:
\begin{eqnarray}
v_s &=& \left\{
\begin{array}{ll}
v_i & (t_{age}<t_1;\ {\rm Free\ expansion\ phase}), \\
v_i \left(\frac{t_{age}}{t_1}\right)^{-3/5} & (t_1 < t_{age} < t_2;\
{\rm Sedov\ phase}), \\
v_i\left(\frac{t_2}{t_1}\right)^{-3/5}\left(\frac{t_{age}}{t_2}\right)^{-2/3}
& (t_2<t_{age};\ {\rm Radiative\ phase}), \\
\label{eq:velocity}
\end{array}
\right.\\
t_1 &=& 2.1\times 10^{2}(E_{51}/n_0)^{1/3}v_{i,9}{}^{-5/3}\ {\rm yr},\\
t_2 &=& 4\times 10^{4}E_{51}{}^{4/17}n_0{}^{-9/17}\ {\rm yr},
\end{eqnarray}
where $v_i = v_{i,9}\times 10^9~{\rm cm~s^{-1}}$,
$E_{\rm ej} = E_{51}\times 10^{51}$~erg,
and $n_0$ are the initial velocity,
initial energy of ejecta,
and the upstream number density, respectively
\citep[e.g.,][]{blondin1998,yamazaki2006}.
%
The model assumes that
the SNR forward shock propagates into a homogeneous medium.
This is, however, not always in the case,
since the SNRs could be located inside of superbubbles
made by pre-explosion stellar winds.
More detailed models
for core-collapsed type SNRs, such as RCW~86
\citep[e.g.,][]{vink1997},
should be applied.
But it is left for future work.

As illustrated above, the X-ray luminosity has little dependence on the
magnetic field. Nevertheless, we  model the evolution of the 
downstream magnetic field, $B_d$, as follows.
%
Usually, the amplified magnetic field $B_{\rm amp}$
is simply assumed to scale with $v_s$  \citep{volk2005,vink2008},
because magnetic field evolution remains ill-understood.
Here we adopt an assumption similar to \citet{volk2005},
leading to a dependence of $B_d\propto v_s$. 
Let the energy density of the amplified magnetic field 
$U_B = B_{\rm amp}{}^2/8\pi$ be proportional to the thermal energy density $U_{th}$,
\[
\frac{B_{\rm amp}{}^2}{8\pi} = \epsilon_BU_{th} = \frac{3}{2}\epsilon_Brn_0kT_d\ \ ,
\]
where $\epsilon_B$, $r$, and $kT_d$ are
the energy-partition parameter, a compression ratio, 
and a downstream temperature
obtained by the Rankine-Hugoniot equation
$kT_d = [(r-1)/{r^2}]m_pv_s{}^2$, respectively.
Then, we obtain
$B_{\rm amp} = [12\pi m_p(r-1)\epsilon_B n_0/r]^{1/2} v_s$.
The downstream magnetic field is thus given by
$B_d={\max}\{B_{\rm amp},rB_{\rm ISM}\}$,
where $B_{\rm ISM}=10~\mu$G is the field strength in the interstellar matter.


For given  $t_{\rm age}$, 
we calculate $v_s$ according to Eq.~(\ref{eq:velocity}),
the shock radius $R_s$ by the integration of 
$v_s$, and the luminosity of synchrotron X-rays numerically
by using the value of $B_d$ and the electron distribution 
given by Eqs.~(\ref{eq:Emax_e}) and (\ref{eq:N_e}).
We adopt $E_{51} = v_{i,9} = 1$, $p=2.0$, $r=4$, $\xi=1$,
and $\epsilon_B = 0.01$ 
as  fiducial parameters.
Note that for $\epsilon_B\sim0.01$, the evolution of the
amplified magnetic field in young
SNRs is approximately reproduced \citep{volk2005}.
This is also consistent with previous observational implications for SNRs
\citep{ghavamian2002,bamba2003,parizot2006}.
One finds that with the above assumption for the magnetic field,
the cooling time of synchrotron X-ray-emitting electrons
is always smaller than the SNR age, $t_{\rm age}$,
so that the X-ray luminosity is well approximated
by Eq.~(\ref{eq:Luminoosity}).
Also note that the value of $\xi$ is near unity,
implying that acceleration is near the Bohm limit.
We assume that the normalization factor of the electron distribution
$A$ is constant with time.
This is justified if the amount of accelerated particles is
proportional to the product of the fluid ram pressure and 
the SNR volume, that is,
$A\propto (n_0v_s{}^2)R_s{}^3$.
In the Sedov phase ($R_s\propto t^{2/5}$)
during which  $\nu_{\rm roll}$ crosses the X-ray band, 
we find that $A$ is constant with time.
%

Fig.~\ref{fig:r_Lx}  shows the results
for $n_0=5.0$, 1.0, and 0.1~cm$^{-3}$.
For the  first two cases,
we set $A=9.0\times 10^{45}(n_0/1~{\rm cm}^{-3})$,
whereas $A=9.0\times 10^{45}$
for $n_0=0.1$~cm$^{-3}$.
$A$ is expressed in the cgs units.
One can see that our model roughly reproduces
the observed trend with reasonable model parameters.
The assumed density, $n_0 = 0.1-5.0~{\rm cm}^{-3}$,
is typical for the interstellar medium around SNRs,
although there are discrepancies for some individual samples.
Some SNRs are likely to be located in bubbles or very low density regions.
For example,
\citet{vink1997} showed that RCW~86 is in a super-bubble
and expands into extremely low density materials.
As a result, this SNR should have a very large radius
compared with its age.
$A$ is also a free parameter
which may change from source to source.
Adjusting $A$ to within 1~order of magnitude,
it is possible to find a solution for the observed density
in each case.
Since our aim is simply to reproduce 
the overall evolutionary trend in an approximate manner,
we ignore such fluctuations in this work.

\section{Discussion}
\label{sec:discuss}

We have reproduced the observed fast drop-off in
synchrotron X-ray luminosity with SNR radius using a quite simple model.
Recently, \citet{patnaude2010} discovered
a decline of a few percent in the synchrotron X-ray power from one source,
Cas~A,
and reached a similar conclusion.

The total and maximum energy
of accelerated particles can be derived from our model.
With the adopted values of $A$ and Eq.~(\ref{eq:N_e}),
we can calculate the total energy of accelerated electrons
with the energy above $m_ec^2$
to be  $10^{47}$--$10^{48}$~ergs,
which is reasonable since 
this range is much smaller than $E_{\rm ej}$.
Note that these estimated values could increase by 1 order of magnitude
if we increase $\xi$ by up to 10.
Even if we alter the value of $p$, the results remain similar
if we appropriately reset the normalization $A$.
This is expected from equation~(\ref{eq:Luminoosity}).
For example, the results for $p=2.2$ are almost perfectly
in agreement with the case of $p=2.0$ if we choose
$A=1.7\times 10^{46}(n_0/1~{\rm cm}^{-3})$
$n_0=5.0$ and 1.0~cm$^{-3}$
and $A=1.7\times 10^{46}$
for $n_0=0.1$~cm$^{-3}$.
In these cases, the total energy of accelerated electrons
with energy above $m_ec^2$ is again
$10^{47}$--$10^{48}$~ergs.
\citet{bamba2003} estimated the energy of injected accelerated electrons
in SN~1006 for the first time.
This was found to be $\sim 10^{44}$--$10^{45}$~ergs
per small segments of filaments,
although with rather large uncertainty.
The total energy of accelerated electrons
is approximately $10^{46}$--$10^{48}$~ergs
considering the size of these segments,
which is consistent with our result.
\citet{bamba2005} show that the enegy injected into accelerated electrons
is always similar in young SNRs,
which is also consistent with our results.
Theoretically, \citet{berezhko2006} estimated the energy injected into
accelerated protons to be $10^{50}$~ergs 
in the case of RX~J1713$-$3946.
This is also consistent with our results
when we consider the electron and proton ratio of $10^{-4}$
in cosmic rays.

\begin{figure}
\epsscale{0.7}
\plotone{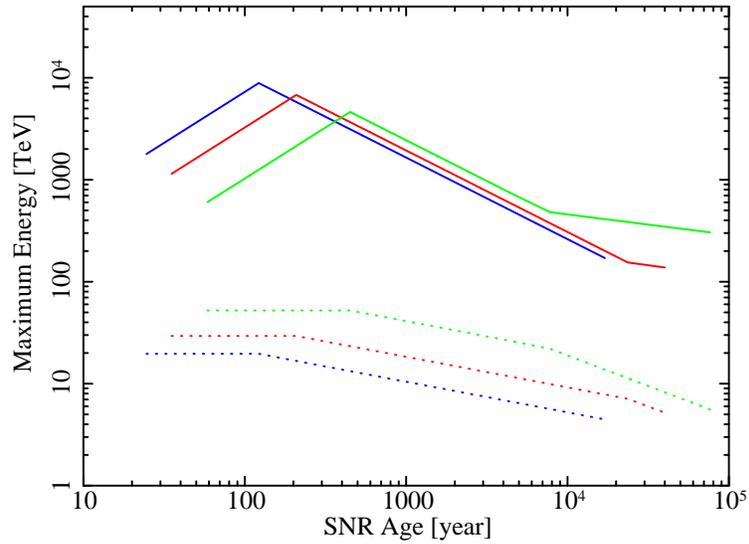}
\caption{The maximum energies
of accelerated protons (solid) and electrons (dotted)
as functions of the SNR age.
The colors match those in Fig.~\ref{fig:r_Lx}.}
\label{fig:Emax_evolution}
\end{figure}

At present, no direct information on the accelerated protons is 
obtained by X-ray observations.
One might consider whether we could estimate
the total energy of accelerated protons
under the assumption of a similar $A$ as for electrons.
However, this is not possible because the injection efficiency is
different in each case.
On the other hand,
the maximum energy of accelerated protons, $E_{p}{}^{max}$,
can be estimated according to our  model.
The synchrotron and $\pi^0$-decay loss timescales are negligible
for protons,
and thus $E_{p}{}^{max}$ is determined by the condition
$t_{\rm acc}=t_{\rm age}$, and we obtain
\begin{equation}
E_{p}{}^{max}(t_{\rm age}) = \frac{4.8\times 10^2}{\xi}
\left(\frac{v_s}{10^9 {\rm cm~s^{-1}}}\right)^2
\left(\frac{B_d}{10~\mu{\rm G}}\right)
\left(\frac{t_{\rm age}}{10^3~{\rm yr}}\right)\ \ {\rm TeV}\ .
\end{equation}
Although the effect of the wave damping and/or escape may be significant
in older systems
\citep[e.g.,][]{zirakashvilli2007,ohira2010}, 
we ignore it here for simplicity.
Figure~\ref{fig:Emax_evolution} shows
$E_{p}{}^{max}$ as a function of $t_{\rm age}$.
In the free expansion phase, 
$E_{p}{}^{max}$ increases in proportion to the age,
and protons are accelerated quickly up to around the knee energy
in $\sim$100~yr.
When the shock velocity decreases,
$E_{p}{}^{max}$ becomes smaller.
Hence $E_{p}{}^{max}$ peaks at $t_{\rm age}=t_1$.
%
%
In contrast to the case of electron acceleration,
$E_{p}{}^{max}(t_1)$
(that is the maximum of $E_{p}{}^{max}$) is larger
for higher upstream density $n_0$,
and it changes only within a factor of $\sim2$ 
for $n_0$ = 0.1--5.0~cm$^{-3}$.
Since  $B_d \propto n_0{}^{1/2}$ and 
$t_1 \propto n_0{}^{-1/3}$, one finds
$E_{p}{}^{max}(t_1)\propto B_dt_1\propto n_0{}^{1/6}$.
Protons are accelerated up to $E_{p}{}^{max}(t_1)\sim 10^{15-16}$~eV
in any SNR, with little density dependence.
This result might be important for explaining the break at
$\sim 10^{15.5}$~eV in the observed cosmic-ray spectrum
with the shock acceleration on SNRs.
On the other hand, one can find 
$E_{e}{}^{max}\propto B_d{}^{-1/2}v_s\propto v_s{}^{1/2}$
if $B_d=B_{\rm amp}\propto v_s$.
Hence the maximum energy of electrons scales as
$E_{e}{}^{max}\propto t^0$ and $\propto t^{-3/10}$
for the free expansion and Sedov phases, respectively,
so that it depends only weakly on the SNR age.

Recent {\it Fermi} observations of middle-aged SNRs have shown the
presence of a gamma-ray spectral break around energies of a few GeV 
\citep[e.g.,][]{abdo2009,abdo2010}.
Several interpretations
have been given \citep[e.g.,][]{inoue2010,li2010,li2011,ohira2011},
 one of which is the escape
of particles from acceleration sites. In this scenario, 
accelerated particles with energies of more than  about 10-100~GeV
have already escaped from the acceleration region \citep{ohira2011}.
In contrast, our present model has shown that even for 
$t_{\rm age}\gtrsim10^4$~yr, $\sim10$~TeV electrons still exist around the
shock front (see Figure~\ref{fig:Emax_evolution}). 
This apparent discrepancy may come from the fact that
most SNRs observed by {\it Fermi} are interacting with molecular clouds,
whereas in our sample there is only one (W~28)
undergoing such a collision.
Hence molecular clouds may play an important role
in dispersing cosmic rays from the SNR shock into the interstellar matter.

\acknowledgments

We would like to thank the anonymous referee for the fruitful comments.
We also thank
S.~Wagner and G.~P{\" u}lhofer
for the analysis of individual sources.
We also thank P.~Gandhi for the English correction.
This work was supported in part by
Grant-in-Aid for Scientific Research
of the Japanese Ministry of Education, Culture, Sports, Science
and Technology (MEXT) of Japan, No.~22684012 (A.~B.),
No.~21740184 and 21540259 (R.~Y.).
and No.~18071001, 22244030, and 21111006 (K.~K.).
K.K. was partly supported by the Center for the Promotion of
Integrated Sciences (CPIS) of
Sokendai.

\begin{deluxetable}{p{6pc}cccccc}
\tabletypesize{\scriptsize}
\tablecaption{Observational journals and physical parameters
for SNRs with synchrotron X-rays.
\label{tab:sample}}
\tablewidth{0pt}
\tablehead{
\colhead{Target} & \colhead{Distance} & \colhead{Size} & \colhead{Radius} & 
\colhead{$L_X$\tablenotemark{a}} &
\colhead{References} \\
 & (kpc) & (arcmin) & (pc) & ($10^{34}$~ergs~s$^{-1}$) &
}
\startdata
G1.9+0.3\dotfill & 8.5 & 1.5 & 1.9 & $4.9_{-0.2}^{+0.1}$ &
(1) \\
Cas~A\dotfill & $3.4_{-0.1}^{+0.3}$ & 5 & $2.5_{-0.1}^{+0.2}$ & $110_{-6}^{+20}$ & 
(2) (3) (4) \\
Tycho\dotfill & $3.8_{-1.1}^{+1.5}$ & 8 & $4.4_{-1.3}^{+1.8}$ & $12_{-6}^{+12}$ & 
(5) (6) (7) \\
RX~J1713$-$3946\dotfill & 1.0 & $65\times55$ & $8.7_{-0.7}^{+0.8}$ & 6.5 & 
(8) (9) \\
SN~1006\dotfill & $2.2\pm0.1$ & 30 & $9.5\pm0.3$ & $2.1_{-0.1}^{+0.2}$ & 
(10) (11) (12) \\
G28.6$-$0.1\dotfill & $7.0_{-1.0}^{+1.5}$ & $13\times9$ & $11\pm2$ & $0.88_{-0.23}^{+0.42}$ & 
(13) \\
Vela Jr.\dotfill & $0.65\pm0.35$ & 120 & $11\pm6$ & $0.42_{-0.33}^{+0.58}$ & 
(14) \\
W28\dotfill & $1.9\pm0.3$ & 48 & $13\pm2$ & $0.069_{-0.02}^{+1.3}$\tablenotemark{b} & 
(15) (16) \\
CTB~37B\dotfill & $10.2\pm3.5$ & 10\tablenotemark{c} & $15\pm5$ & $0.97_{-0.59}^{+3.1}$ & 
(17) (18) \\
G32.45+0.1\dotfill & $17_{-4}^{+7}$ & 6 & $15_{-3}^{+6}$ & $4.5_{-1.9}^{+4.5}$ & 
(19) \\
G156.2+5.7\dotfill & $1.0_{-0.7}^{+0.3}$ & 110 & $16_{-11}^{+5}$ & $0.060_{-0.055}^{+0.43}$\tablenotemark{b} & 
(20) (21) (22) \\
RCW~86\dotfill & $2.8\pm0.4$ & 42 & $17\pm2$ & $3.8_{-1.0}^{+1.2}$ & 
(23) (24) (25) 
\enddata
\tablecomments{(1) \citet{reynolds2008};
(2) \citet{reed1995}; (3) \citet{helder2008};
(4) \citet{lazendic2006}; 
(5) \citet{krause2008}; (6) \citet{tamagawa2009};
(7) \citet{seward1983}; 
(8) \citet{fukui2003}; (9) \citet{slane1999};
(10) \citet{winkler2003}; (11) \citet{bamba2008};
(12) \citet{yamaguchi2008};
(13) \citet{bamba2001}; 
(14) \citet{aharonian2007};
(15) \citet{velazquez2002};
(16) Nakamura et al., submitted; 
(17) \citet{caswell1975};
(18) \citet{nakamura2009};
(19) \citet{yamaguchi2004};
(20) \citet{gerardy2007}; (21) \citet{yamauchi1999}; (22) \citet{katsuda2009};
(23) \citet{rosado1996}; (24) \citet{bamba2000};
(25) \citet{vink2006}
}
\tablenotetext{a}{In the 2--10~keV band.}
\tablenotetext{b}{The upper-bound of X-ray luminosity is calculated
with the assumption that the SNR uniformly emits synchrotron X-rays,
since present observations do not cover the entire remnant.}
\tablenotetext{c}{The size of radio partial shell is used.}
\end{deluxetable}

\end{document}